\documentstyle[11pt]{article}

\begin{document}

\title{{\Large {\bf Self-similar Bianchi models I: Class A models}}}
\author{Pantelis S. Apostolopoulos \\
%EndAName
{\small {\it Department of Physics, Section of
Astrophysics-Astronomy-Mechanics},}\\
{\small {\it University of Athens, Panepistemiopolis 157 83, Athens, GREECE}}}
\maketitle

\begin{abstract}
We present a study of Bianchi class A tilted cosmological models admitting a
proper homothetic vector field together with the restrictions, both at the
geometrical and dynamical level, imposed by the existence of the simply
transitive similarity group. The general solution of the symmetry equations
and the form of the homothetic vector field are given in terms of a set of
arbitrary integration constants. We apply the geometrical results for tilted
perfect fluids sources and give the general Bianchi II self-similar solution
and the form of the similarity vector field. In addition we show that
self-similar perfect fluid Bianchi VII$_0$ models and irrotational Bianchi VI%
$_0$ models do not exist.

{\bf PACS: 4.20.-q, 4.20.Ha}
\end{abstract}

\section{Introduction}

\setcounter{equation}{0}

The simplest generalisation of the Friedmann-Lema\^{i}tre universes are the
Spatially Homogeneous (SH) models which admit a local group $G_r$ ($r=3,4$)
of isometries acting on 3-dimensional spacelike hypersurfaces. From the
physical point of view they serve as the basis in the study of anisotropies
at early times of the universe evolution and play a key role in the
understanding of the underlying geometrical/physical properties and insights
of more general cosmological solutions of the Field Equations (FE): 
\begin{equation}
R_{ab}=T_{ab}-\frac T2g_{ab}.  \label{sx1.1}
\end{equation}
An extensively studied subclass of SH models are the Bianchi models where $%
r=3$ which can be divided further into two major subclasses: orthogonal
Bianchi cosmologies where the fluid four-velocity $n^a$ is orthogonal to the
group orbits \cite{Ellis-MacCallum} leading to the {\em non-tilted} models
and Bianchi cosmologies where the normalised fluid velocity $u^a$ is not
aligned with $n^a$ representing {\em tilted} models \cite{King-Ellis}.

Although in SH models, the FE are reduced to a system of ordinary
differential equations, not many exact solutions are known, especially in
the case $r=3$. This has initiated the examination of their dynamics using
more sophisticated and predictive methods: the FE are reformulated as an
autonomous system and using the methods of the theory of dynamical systems,
the evolution of a specific model is studied in the so called {\em state
space} which represents the set of all the physical states (at some instant
of time) of the corresponding model \cite{Wainwright-Ellis,Carr-Coley}. The
main idea is to identify the evolution of a specific model with an orbit
(solution curve) in state space and examine its geometrical-with direct
physical interpretation-properties. One of the striking results of this
approach is that {\em transitively self-similar} SH models are important in
describing the asymptotic (i.e. at early and late times) states of the
evolution for more general SH models. In particular, self-similar SH models
arise as equilibrium points which determine various stable and unstable
invariant submanifolds of the state space of the evolution equations. The
determination of these invariant submanifolds allows to gain deeper insight
into the intermediate evolution of the SH cosmological models. Furthermore
there was a strong evidence that any SH model could be approximated by
self-similar SH models in the asymptotic regimes i.e. near the initial
cosmological singularity and at late times. However it has been proved
recently \cite{Wainwright-Hancock-Uggla,Apostol-Tsampa7} that Bianchi VIII
and IX models are not (asymptotically) self-similar providing a
counterexample in this conjecture. Nevertheless Bianchi type IX cosmologies
are successively approximated (at the asymptotic regimes) by an infinite
sequence of self-similar models (Kasner vacuum solutions) reinforce the
belief that self-similar models have a significant role in the structure of
the SH dynamical state space.

It becomes clear that self-similar Bianchi models are of primary
cosmological importance. We will focus attention on Bianchi models ($r=3$)
admitting a proper homothetic vector field (HVF) ${\bf H}$ acting {\em %
simply transitively} on space-time: 
\begin{equation}
{\cal L}_{{\bf H}}g_{ab}=2\psi g_{ab}  \label{sx1.2}
\end{equation}
where $\psi =$const. is the homothetic factor which essentially represents
the (constant) scale transformation of the geometrical and dynamical
variables.

The main goal of this paper is to provide the complete set of self-similar
Bianchi models of class A (in particular for the "missing" types II,VI$_0$%
,VII$_0$) {\em irrespective} the matter source of the gravitational field.
In the case of non-vacuum fluid models the four-velocity subjects only to
the assumption that is mapped conformally by the 4-dimensional similarity
group. We utilize the purely geometrical results and study the case of
tilted perfect fluid models. Due to the intrinsic complexity of the tilted
models, up to date, not all the self-similar models have been determined
adding to the difficulty of qualitatively analysing Bianchi models.
Therefore the self-similar metrics presenting in this paper can be used to
determine the general self-similar tilted perfect fluid Bianchi models of
class A (note that in the vacuum and non-tilted case considerable
information are available \cite{Hsu-Wainwright}).

An outline of the paper is as follows: section 2 is devoted to a brief
presentation of some of the basic properties of Bianchi models using the
metric approach in which the basic variables are the frame components of the
metric w.r.t. the canonical 1-forms \cite{Ryan-Shepley1,MacCallum2}. We then
introduce a 1+3 decomposition of the tilted fluid velocity $u^a$ and show
that the frame components of $u^a$ satisfy restrictions imposed by the
structure of the isometry algebra of the SH models. In section 3 we confine
our study to Bianchi class A models and present a set of constraints coming
from the existence of a proper HVF. These constraints are used to determine
explicitly the frame components of the metric, the fluid velocity and the
corresponding similarity vector fields. In section 4 we discuss the physical
implications of these, purely geometrical, results and examine the existence
of tilted perfect fluid models. For Bianchi II models we regain the {\em %
general} self-similar tilted perfect fluid solution and prove that {\em %
there are no irrotational self-similar tilted Bianchi VI}$_0${\em \ models}.
We also show that there {\em no self-similar tilted perfect fluid Bianchi VII%
}$_0${\em \ models}. We conclude in section 5 discussing our results and
further avenues of research.

\section{Preliminaries}

\setcounter{equation}{0}

In this section we briefly review some basic features of the Bianchi models
regarding the effects, both at the level of Geometry and Physics, which
arise from the existence of the 3-parameter group of isometries. Since the
theory is well known (see e.g. \cite{Wainwright-Ellis} and \cite{Jantzen1}
and references cited therein) we present hear only those results which will
be useful in the subsequent sections. Following Ellis et al. \cite
{Wainwright-Ellis} we use the so called {\em metric approach} in which the
basic variables are the frame components of the metric with respect to the
set of canonical 1-forms. Although alternative techniques provide powerful
unified formalisms of qualitatively analysing Bianchi models, the metric
approach has the advantage of dealing with coordinate components hence we
can determine the explicit (local) form of the self-similar metrics and
consequently to study the resulting models directly. Throughout the paper
the index convention is such that latin indices $a,b,c,...$are tensor
components and take the values $0,1,2,3$ whereas Greek indices $\alpha
,\beta ,\gamma ,...=1,2,3$ denotes frame components (non-tensorial) of the
corresponding quantities. A semicolon ($;$) denotes covariant
differentiation and a comma ($,$) denotes partial differentiation w.r.t.
following index coordinate.

\subsection{Geometry}

Bianchi models are characterized by the existence of a $G_3$ Lie group of
isometries acting on 3-dimensional spacelike orbits ${\cal C}$ with
generators the Killing vector fields (KVFs) ${\bf X}_\alpha $ tangent to the
group orbits. The spacelike hypersurfaces ${\cal C}$ and the associated
group $G_3$ of isometries uniquely determine a unit timelike congruence $n^a$
($n^an_a=-1$) normal to the spatial foliations ${\cal C}$ which satisfies:

\begin{equation}
n_{[a;b]}=0=n_{a;b}n^b\Leftrightarrow \frac 12{\cal L}_{{\bf n}}g_{ab}\equiv
n_{a;b}=\sigma _{ab}+\frac \theta 3h_{ab}  \label{sx2.4}
\end{equation}
where $\sigma _{ab},\theta =n_{a;b}g^{ab},h_{ab}=g_{ab}+n_an_b$ are the
trace-free symmetric shear tensor, the expansion and the projection tensor
respectively associated with the timelike congruence $n^a$ \cite{Ellis1}. It
turns out that the (irrotational and geodesic) timelike congruence $n^a$ is
invariant under the action of the group $G_3$ i.e. the KVFs commute with $%
n^a $:

\begin{equation}
\lbrack {\bf X}_\alpha ,{\bf n}]=0.  \label{sx2.5}
\end{equation}
Therefore local coordinates $\{t,x^\alpha \}$ can be chosen such that $%
n_a=-t_{,a}$ and the spacelike foliations ${\cal C}$ coincide with the
hypersurfaces of constant $t$ \cite{Ellis-MacCallum}.

Introducing a set of invariant basis of vector fields ${\bf e}_\alpha $ and
its dual ${\bf \omega }^\beta $ such that: 
\begin{equation}
\lbrack {\bf X}_\alpha ,{\bf e}_\beta ]=0\qquad [{\bf X}_\alpha ,{\bf \omega 
}^\beta ]=0  \label{sx2.6}
\end{equation}
the full four dimensional metric can be written: 
\begin{equation}
ds^2=-dt^2+g_{\alpha \beta }(t){\bf \omega }^\alpha {\bf \omega }^\beta
\label{sx2.8}
\end{equation}
and we have ensured the constancy of the 3-metric $g_{\alpha \beta }$ on
each orbit ${\cal C}$.

In addition the invariant basis ${\bf e}_\alpha $ and its dual ${\bf \omega }%
^\alpha $ satisfy the corresponding commutation relations:

\begin{equation}
\lbrack {\bf e}_\alpha ,{\bf e}_\beta ]=-C_{\alpha \beta }^\gamma {\bf e}%
_\gamma \qquad ,\qquad d{\bf \omega }^\gamma =\frac 12C_{\alpha \beta
}^\gamma {\bf \omega }^\alpha \wedge {\bf \omega }^\beta  \label{sx2.9}
\end{equation}
where $d$ stands for the usual exterior derivative of 1-forms, $\wedge $
denotes the usual exterior product\footnote{%
Note that the sign in (\ref{sx2.9}) depends on the choice of the initial
condition of the differential equations (\ref{sx2.6}).} and $C_{\alpha \beta
}^\gamma =-C_{\beta \alpha }^\gamma $ are the structure constants of the
group.

\subsection{Physics}

As we have mentioned in the Introduction, attention is focused on perfect
fluid sources which can be described in terms of the normalised
four-velocity of the fluid $u^a$ ($u^au_a=-1$): 
\begin{equation}
T_{ab}=(\tilde{\mu}+\tilde{p})u_au_b+\tilde{p}g_{ab}  \label{sx2.10}
\end{equation}
where $\tilde{\mu},\tilde{p}$ are the energy density and the pressure
measured by the observers comoving with the fluid velocity $u^a$. Using the
FE we may express the Ricci tensor in a similar form:

\begin{equation}
R_{ab}=(\tilde{\mu}+\tilde{p})u_au_b+\frac{(\tilde{\mu}-\tilde{p})}2g_{ab}.
\label{sx2.11}
\end{equation}
A straightforward consequence of the spatial homogeneity is that there is
only one essential dynamical coordinate $t$. This follows from the
invariance of the matter tensor under the action of the $G_3$ \cite
{Coley-Tupper1}: 
\begin{equation}
{\cal L}_{{\bf X}_\alpha }T_{ab}={\cal L}_{{\bf X}_\alpha }R_{ab}=0,[{\bf X}%
_\alpha ,{\bf u}]=0,{\bf X}_\alpha (\tilde{\mu})={\bf X}_\alpha (\tilde{p}%
)=0.  \label{sx2.12}
\end{equation}
We note that the last two equations imply that perfect fluid models
necessarily have a barotropic equation of state $\tilde{p}=\tilde{p}(\tilde{%
\mu})$. Furthermore and for the purposes of the present paper, it is
convenient to decompose $u^a$ parallel and perpendicular to $n^a$ as follows:

\begin{equation}
{\bf u}=\Gamma {\bf n}+\Gamma \Delta _\alpha {\bf \omega }^\alpha =\Gamma 
{\bf n+}\Gamma B^\alpha {\bf e}_\alpha  \label{sx2.13}
\end{equation}
where $B^\alpha (t),\Delta _\alpha (t)$ are the frame components of the
four-velocity $u^a$ and $\Gamma $ is a smooth function of the time
coordinate $t$ satisfying the constraint: 
\begin{equation}
\Gamma =\left( 1-B^\alpha \Delta _\alpha \right) ^{-\frac 12}  \label{sx2.14}
\end{equation}
and $B^\alpha \Delta _\alpha <1$.

The fluid velocity $u^a$, although invariant under the action of the group
of isometries, {\em is not} necessarily parallel to the normal timelike
congruence $n^a$, leading to {\em tilted} perfect fluid models i.e. the
fluid flow lines are not orthogonal to the surfaces of spatial homogeneity
(the quantity $B^\alpha \Delta _\alpha $ of the projection of $u^a$ is
directly connected with the so called ''{\em tilt direction}'' \cite
{King-Ellis}). In order to establish the relation between the dynamical
quantities defined by the timelike vector fields $n^a$,$u^a$, we consider
the 1+3 decomposition of the matter tensor induced by the normal timelike
vector field $n^a$ \cite{Ellis1}: 
\begin{equation}
T_{ab}=\mu n_an_b+ph_{ab}+2q_{(a}n_{b)}+\pi _{ab}  \label{sx2.15}
\end{equation}
where

\begin{equation}
\mu =T_{ab}n^an^b,\mbox{ }p=\frac 13T_{ab}(n^an^b+g^{ab}),\mbox{ }%
q_a=-h_a^cT_{cd}n^d,\pi _{ab}=h_a^ch_b^dT_{cd}-\frac 13(h^{cd}T_{cd})h_{ab}
\label{sx2.16}
\end{equation}
are the dynamical quantities relative to $n^a$.

Using equations (\ref{sx2.10}) and (\ref{sx2.15}) we obtain the relations 
\cite{van Elst-Uggla}:

\begin{equation}
\mu =\tilde{\mu}+\Gamma ^2v^2(\tilde{\mu}+\tilde{p})\quad ,\quad p=\tilde{p}%
+\frac 13\Gamma ^2v^2(\tilde{\mu}+\tilde{p})  \label{sx2.17}
\end{equation}
\begin{equation}
q_a=\Gamma ^2(\tilde{\mu}+\tilde{p})Y_a\quad ,\quad \pi _{ab}=\Gamma ^2(%
\tilde{\mu}+\tilde{p})\left[ Y_aY_b-\frac 13Y^2\left( g_{ab}+n_an_b\right)
\right]  \label{sx2.19}
\end{equation}
where 
\begin{equation}
{\bf Y}=B^\alpha {\bf e}_\alpha =\Delta _\alpha {\bf \omega }^\alpha \qquad
,Y^2=Y^aY_a=B^\alpha \Delta _\alpha .  \label{sx2.21}
\end{equation}
Equations (\ref{sx2.19}) indicate that the energy flux vector field $q_a$
and the trace-free anisotropic pressure tensor $\pi _{ab}$ are completely
determined by the energy density and the pressure of the fluid and the
spatial projection $Y^a$ of $u^a$ to the surfaces of spatial homogeneity.

The system of equations is supplemented with the conservation equation:

\begin{equation}
T_{;b}^{ab}=0  \label{sx2.22}
\end{equation}
which for the case of the tilted perfect fluid source is reduced to:

\begin{equation}
\tilde{\mu}_{;a}u^a=-(\tilde{\mu}+\tilde{p})\tilde{\theta}  \label{sx2.23}
\end{equation}
\begin{equation}
\tilde{h}_a^k\tilde{p}_{;k}=-(\tilde{\mu}+\tilde{p})\dot{u}_a  \label{sx2.24}
\end{equation}
and we have used the 1+3 decomposition of the first derivatives of the
four-velocity \cite{Ellis1}: 
\begin{equation}
u_{a;b}=\tilde{\sigma}_{ab}+\frac{\tilde{\theta}}3\tilde{h}_{ab}+\tilde{%
\Omega}_{ab}-\dot{u}_{[a}u_{b]}  \label{sx2.25}
\end{equation}
where $\tilde{\sigma}_{ab},\tilde{\theta}=u_{a;b}g^{ab},\tilde{\Omega}_{ab},%
\dot{u}_a$ are the corresponding kinematical quantities of the fluid
velocity $u^a$, namely, the trace-free shear tensor, expansion, rotation
tensor and acceleration respectively and $\tilde{h}_{ab}$ is the projection
tensor associated with $u^a$. From (\ref{sx2.25}) we obtain easily: 
\begin{equation}
u_{[a;b]}=\tilde{\Omega}_{ab}-\dot{u}_{[a}u_{b]}.  \label{sx2.26}
\end{equation}
\begin{equation}
\frac 12{\cal L}_{{\bf u}}g_{ab}\equiv u_{(a;b)}=\tilde{\sigma}_{ab}+\frac{%
\tilde{\theta}}3\tilde{h}_{ab}-\dot{u}_{(a}u_{b)}.  \label{sx2.27}
\end{equation}
Due to spatial homogeneity (third and fourth of equations (\ref{sx2.12}))
all the (kinematical and dynamical) quantities are invariant under the $G_3$
(which is equivalent with the vanishing of their Lie derivative with respect
to the KVFs).

The decomposition (\ref{sx2.13}) implies:

\begin{equation}
u_{a;b}=\Gamma _{,b}n_a{\bf +}\Gamma n_{a;b}+\Gamma _{,b}\Delta _\alpha
\omega _a^\alpha +\Gamma \left( \Delta _\alpha \right) _{,b}\omega _a^\alpha
+\Gamma \Delta _\alpha \omega _{a;b}^\alpha .  \label{sx2.28}
\end{equation}
We note that a similar relation holds if we apply the decomposition with
respect to the invariant basis ${\bf e}_\alpha $. The skew symmetric part of
equation (\ref{sx2.28}) gives: 
\begin{equation}
u_{[a;b]}=-\left( \Gamma \Delta _\alpha \right) ^{\bullet }n_{[b}\omega
_{a]}^\alpha +\Gamma \Delta _\alpha \omega _{[a;b]}^\alpha  \label{sx2.29}
\end{equation}
where a dot denotes differentiation w.r.t. $t$.

Contracting equation (\ref{sx2.29}) with $e_\alpha ^ae_\beta ^b$ and using (%
\ref{sx2.9}) we obtain:

\begin{equation}
u_{[a;b]}e_\alpha ^ae_\beta ^b=\frac 12\Gamma C_{\alpha \beta }^\gamma
\Delta _\gamma .  \label{sx2.30}
\end{equation}
In the case of a linear equation of state $\tilde{p}=(\gamma -1)\tilde{\mu}$
the conservation law (\ref{sx2.24}) and equation (\ref{sx2.26}) give:

\begin{equation}
u_{[a;b]}e_\alpha ^ae_\beta ^b=\tilde{\Omega}_{ab}e_\alpha ^ae_\beta ^b
\label{sx2.31}
\end{equation}
\begin{equation}
\dot{u}_{[a}u_{b]}e_\alpha ^ae_\beta ^b=0.  \label{sx2.32}
\end{equation}
Equation (\ref{sx2.30}) relates the frame (spatial) components of $u_{[a;b]}$
and $\tilde{\Omega}_{ab}$ with the frame components $\Delta _\alpha $ of the
four-velocity $u^a$ and the structure constants of the Lie group $G_3$ (we
note that a rather similar equation has been proved in \cite{King-Ellis}
based on the Ellis's orthonormal tetrad formalism). Therefore a simple
inspection of the group structure of the spatial homogeneous models, jointly
with the FE, show that the kinematical and dynamical variables of the
Bianchi models depend directly on the group structure. For example it is
straightforward to prove that there are no tilted perfect fluid Bianchi I
models whereas Bianchi types VIII and IX must necessarily have non-zero
vorticity \cite{King-Ellis}. Indeed in type I the structure constants all
vanish thus $\tilde{\Omega}_{ab}=0\Leftrightarrow u_{[a;b]}=-\dot{u}%
_{[a}u_{b]}$ which, by means of equation (\ref{sx2.29}) and the evolution
equation of $\tilde{\Omega}_{ab}$ implies $\dot{u}_an^a=0$. Then, provided
that $\tilde {\mu} _{;a}\neq 0$, the conservation law (\ref{sx2.24}) leads
to $\dot{u}_a=0$ and the fluid is non tilted. In types VIII and IX, assuming
that $\tilde{\Omega}_{ab}=0$ equation (\ref{sx2.30}) implies that $\Delta
_\gamma =0$ and the fluid is also non-tilted (and may be interpreted as
anisotropic fluid). In the next section we draw attention only on Bianchi
models of class A which possess a simply transitively self-similarity group.
The case of class B will be studied in a forthcoming paper \cite{Apostol8}.

\section{Self-similar Bianchi models of class A}

\setcounter{equation}{0}

A transitively self-similar Bianchi model admits a proper HVF ${\bf H}$
which together with the isometry group, form a simply transitive group of
homotheties $H_4$ acting on four dimensional orbits and has generators $\{%
{\bf H},{\bf X}_\alpha \}$ spanned the associated Lie algebra denoted as $%
{\cal H}_4$. Our main concern in this section will be to find the general
solution of the symmetry equations (\ref{sx1.2}) in Bianchi class A models.
However it is useful to present first the restrictions on the kinematical
and dynamical variables of the Bianchi models induced from the existence of
the HVF. For this reason we shall need the following well known result \cite
{Maartens-Mason-Tsamparlis1}:

{\em Proposition 1}

For {\em every} unit timelike vector field $u^a$ and a HVF $H^a=M_{{\bf H}%
}u^a+w^a$ where $w^au_a=0$ the following identity holds:

\begin{equation}
L_{{\bf H}}u_a=\left[ \left( M_{{\bf H}}\right) _{;k}u^k+\dot{u}_kH^k\right]
u_k+2\omega _{ab}H^b+M_{{\bf H}}\left[ \dot{u}_a-\left( \ln M_{{\bf H}%
}\right) _{;k}h_a^k\right]  \label{sx3.01}
\end{equation}
where $\dot{u}_a,\omega _{ab},h_a^k$ are the acceleration, the rotation and
the projection tensor associated with the unit timelike vector field $u^a$.

\subsection{Geometry}

A direct calculation shows that the commutator of a HVF with a KVF is always
a KVF which implies that the structure constants $C_{ab}^d$ of the
homothetic Lie algebra ${\cal H}_4$ satisfy $C_{ab}^4=0$. Moreover the
Jacobi identities for $C_{ab}^d$ are equivalent to the relations:\ 

\begin{equation}
C_{\delta [a}^\gamma C_{bc]}^\delta =0.  \label{sx3.1}
\end{equation}
Following the considerations of the previous section, we decompose the HVF
parallel and normal to $n^a$:

\begin{equation}
{\bf H}=H{\bf n}+A^\alpha {\bf X}_\alpha  \label{sx3.2}
\end{equation}
where $H$ and $A^\alpha $ are smooth functions of the space-time manifold.

From the commutator relations $[{\bf X}_\alpha ,{\bf H}]=C_{\alpha 4}^\gamma 
{\bf X}_\gamma $ we get:

\begin{equation}
{\bf X}_\alpha (H)=0\qquad ,\qquad {\bf n}(H)=\psi  \label{sx3.3}
\end{equation}

\begin{equation}
{\bf n}(A^\beta )=0\qquad ,\qquad {\bf X}_\alpha (A^\gamma )+A^\beta
C_{\alpha \beta }^\gamma =C_{\alpha 4}^\gamma  \label{sx3.4}
\end{equation}
where we have used (\ref{sx3.01}) and the fact that, since ${\bf n}$ is
geodesic and irrotational, one has $[{\bf H,n]}=-{\bf n}(H){\bf n}+[A^\alpha 
{\bf X}_\alpha ,{\bf n]=}-\psi {\bf n}$ where ${\bf n}(H)=\psi $. Equations (%
\ref{sx3.3}) and (\ref{sx3.4}) determine the functions $A^\gamma $ up to
constants of integration. However some of them can be ignored by means of
appropriate transformations which maintain the form of the metric and the
KVFs. Therefore the HVF is easily determined, provided that $C_{\alpha
4}^\gamma $ and $C_{\alpha \beta }^\gamma $ are given and subsequently,
solving the symmetry equations (\ref{sx1.2}), we determine the frame
components $g_{\alpha \beta }(t)$ of the metric.

\subsection{Physics}

Using the well known relations ${\cal L}_{{\bf H}}R_{ab}={\cal L}_{{\bf H}%
}T_{ab}=0$ \cite{Yano}, it can be shown \cite{Coley-Tupper1,Coley-Tupper2}
that the fluid flow lines are mapped conformally by the HVF and the equation
of state is necessarily linear \cite{Wainwright}. Moreover we deduce that in
a self-similar SH perfect fluid model the ''tilt angle'' is constant and the
projection $Y^a$ of the fluid velocity is also conformally mapped by the
HVF: 
\begin{equation}
{\bf H}(\Gamma )=0\Leftrightarrow \Gamma _{;a}=0\qquad ,\qquad {\cal L}_{%
{\bf H}}Y^a=-\psi Y^a\qquad ,\qquad {\cal L}_{{\bf H}}Y_a=\psi Y_a
\label{sx3.12}
\end{equation}
which by means of equations (\ref{sx2.28}) and (\ref{sx2.29}) imply:

\begin{equation}
u_{a;b}=\Gamma n_{a;b}+\Gamma \left( \Delta _\alpha \right) _{,b}\omega
_a^\alpha +\Gamma \Delta _\alpha \omega _{a;b}^\alpha  \label{sx3.13}
\end{equation}
\begin{equation}
u_{[a;b]}=-\Gamma \dot{\Delta}_\alpha n_{[b}\omega _{a]}^\alpha +\Gamma
\Delta _\alpha \omega _{[a;b]}^\alpha .  \label{sx3.14}
\end{equation}
We may further observe that the frame components $B^\alpha ,\Delta _\alpha $
satisfy: 
\begin{equation}
B^\alpha \Delta _\alpha =const.  \label{sx3.15}
\end{equation}
Equations (\ref{sx2.12}) and (\ref{sx3.12}) can be used to determine, up to
a set arbitrary constants of integration, the form of the fluid velocity.

\subsection{Solution of the symmetry equations}

Solving the system of equations (\ref{sx1.2}), (\ref{sx3.3}), (\ref{sx3.4})
and (\ref{sx3.12}) we can find the explicit expressions of the self-similar
metric (the {\em frame components} of the metric i.e. the functions $%
g_{\alpha \beta }(t)$ in (\ref{sx2.8})), the homothetic vector field and the
tilted four-velocity $u^a$ for Bianchi class A models (we ignore the cases
of Bianchi types I, VIII and IX since no tilted perfect fluid solutions
exist). These results hold for a generic type of matter provided that, for
fluid models, the four-velocity satisfies ${\cal L}_{{\bf H}}u^a=-\psi u^a$.

\underline{{\bf Type II}}

In local coordinates $\{t,x,y,z\}$ and following the notations of \cite
{Ryan-Shepley1}, the KVs $\{{\bf X}_\alpha \}$ and the canonical 1-forms $\{%
{\bf \omega }^\alpha \}$ are:

\begin{equation}
{\bf X}_1=\partial _y\qquad ,{\bf X}_2=\partial _z\qquad ,{\bf X}_3=\partial
_x+z\partial _y.  \label{sx3.20}
\end{equation}
\begin{equation}
{\bf \omega }^1=dy-xdz\qquad ,{\bf \omega }^2=dz\qquad ,{\bf \omega }^3=dx.
\label{sx3.20a}
\end{equation}
The non-vanishing structure constants are $C_{23}^1=1$ and the Jacobi
identities (\ref{sx3.1}) imply that the remaining non-vanishing structure
constants $C_{\beta 4}^\alpha $ are \cite{Petrov,Koutras}:

\begin{equation}
C_{14}^1=a\qquad ,C_{24}^2=1\qquad ,C_{34}^3=a-1  \label{sx3.21}
\end{equation}
\begin{equation}
C_{14}^1=2\qquad ,C_{24}^2=1\qquad ,C_{34}^2=1\qquad ,C_{34}^3=1
\label{sx3.22}
\end{equation}
\begin{equation}
C_{14}^1=a\qquad ,C_{24}^3=1\qquad ,C_{34}^2=-1\qquad ,C_{34}^3=a\qquad
(a^2<4)  \label{sx3.23}
\end{equation}
We consider subcases:

\underline{Case A$_1$}

\noindent {\bf HVF}

\begin{equation}
{\bf H}=\psi t\partial _t+\left( a-1+b\right) \partial _x+ay\partial
_y+z\partial _z  \label{sx3.24}
\end{equation}
{\bf Fluid velocity}

\begin{equation}
\Delta _1=v_1t^{(\psi -a)/\psi }\qquad ,\Delta _2=t^{(\psi -1)/\psi }\left(
v_2+\frac{v_1b\ln t}\psi \right) \qquad ,\Delta _3=v_3t^{(\psi +1-a)/\psi }
\label{sx3.25}
\end{equation}
{\bf Metric}

\begin{equation}
g_{\alpha \beta }=\left( 
\begin{array}{ccc}
c_{11}t^{2(\psi -a)/\psi } & t^{(2\psi -a-1)/\psi }\left[ c_{12}+\frac{%
bc_{11}\left( \ln t\right) }\psi \right] & c_{13}t^{(2\psi -2a+1)/\psi } \\ 
t^{(2\psi -a-1)/\psi }\left[ c_{12}+\frac{bc_{11}\left( \ln t\right) }\psi
\right] & t^{2(\psi -1)/\psi }\left[ c_{22}+\frac{b^2c_{11}\left( \ln
t\right) ^2}{\psi ^2}+\frac{2bc_{12}\ln t}\psi \right] & t^{(2\psi -a)/\psi
}\left[ c_{23}+\frac{bc_{13}\ln t}\psi \right] \\ 
c_{13}t^{(2\psi -2a+1)/\psi } & t^{(2\psi -a)/\psi }\left[ c_{23}+\frac{%
bc_{13}\left( \ln t\right) }\psi \right] & c_{33}t^{2(\psi -a+1)/\psi }
\end{array}
\right)  \label{sx3.26}
\end{equation}
where $b=0$ when $a\neq 1$.

\underline{Case A$_2$}

\noindent {\bf HVF}

\begin{equation}
{\bf H}=\psi t\partial _t+x\partial _x+\frac{x^2+4y}2\partial _y+\left(
x+z\right) \partial _z  \label{sx3.30}
\end{equation}
{\bf Fluid velocity}

\begin{equation}
\Delta _1=v_1t^{(\psi -2)/\psi }\qquad ,\Delta _2=v_2t^{(\psi -1)/\psi
}\qquad ,\Delta _3=t^{(\psi -1)/\psi }\left( v_3-\frac{v_2}\psi \ln t\right)
\label{sx3.31}
\end{equation}
{\bf Metric}

\begin{equation}
g_{\alpha \beta }=\left( 
\begin{array}{ccc}
c_{11}t^{2(\psi -2)/\psi } & c_{12}t^{(2\psi -3)/\psi } & t^{(2\psi -3)/\psi
}\left( c_{13}-\frac{c_{12}\ln t}\psi \right) \\ 
c_{12}t^{(2\psi -3)/\psi } & c_{22}t^{2(\psi -1)/\psi } & t^{2(\psi -1)/\psi
}\left[ c_{23}-\frac{c_{22}\ln t}\psi \right] \\ 
t^{(2\psi -3)/\psi }\left( c_{13}-\frac{c_{12}\ln t}\psi \right) & t^{2(\psi
-1)/\psi }\left[ c_{23}-\frac{c_{22}\ln t}\psi \right] & t^{2(\psi -1)/\psi
}\left[ c_{33}+\frac{c_{22}\left( \ln t\right) ^2}{\psi ^2}-\frac{2c_{23}\ln
t}\psi \right]
\end{array}
\right)  \label{sx3.32}
\end{equation}

\underline{Case A$_3$}

\noindent {\bf HVF}

\begin{equation}
{\bf H}=\frac a{2-p_2}t\partial _t+(ax+z)\partial _x+\left( ay+\frac{z^2-x^2}%
2\right) \partial _y-x\partial _z  \label{sx3.34}
\end{equation}
{\bf Fluid velocity}

\begin{equation}
\Delta _1=v_1t^{p_2-1}\quad ,\Delta _2=t^{p_2/2}\left[ v_{23}\cos \left( 
\frac{p_1\ln t}2\right) +v_{32}\sin \left( \frac{p_1\ln t}2\right) \right]
\quad ,\Delta _3=\frac a{2-p_2}\left[ \Delta _2-t\left( \Delta _2\right)
_{,t}\right]  \label{sx3.34a}
\end{equation}

{\bf Metric}

\begin{equation}
g_{\alpha \beta }=\left( 
\begin{array}{ccc}
c_{11}t^{2(p_2-1)} & g_{21} & g_{31} \\ 
\frac a{2-p_2}\left[ t\left( g_{13}\right) _{,t}-2\left( p_2-1\right)
g_{13}\right] & \frac a{2-p_2}\left[ t\left( g_{23}\right)
_{,t}-p_2g_{23}+g_{33}\right] & g_{32} \\ 
t^{(3p_2-2)/2}\left[ c_{13}\cos \left( \frac{p_1\ln t}2\right) +c_{31}\sin
\left( \frac{p_1\ln t}2\right) \right] & \frac a{2(2-p_2)}\left[ t\left(
g_{33}\right) _{,t}-2\left( p_2-1\right) g_{33}\right] & g_{33}
\end{array}
\right)  \label{sx3.35a}
\end{equation}
where:

\begin{equation}
p_1=\frac{\sqrt{4-a^2}}\psi \quad ,p_2=\frac{2\psi -a}\psi  \label{sx3.35b}
\end{equation}
\begin{equation}
g_{33}=t^{p_2}\left[ c_{33}+c_{32}\cos \left( p_1\ln t\right) +c_{23}\sin
\left( p_1\ln t\right) \right] .  \label{sx3.35c}
\end{equation}
We note that in all cases A$_1$,A$_2$,A$_3$ the quantities $v_\alpha
,c_{\alpha \beta },a,b$ are constants of integration.

\underline{{\bf Type VI}$_0$}

The KVs $\{X_\alpha \}$ and the dual basis $\{{\bf \omega }^\alpha \}$ are:

\begin{equation}
{\bf X}_1=\partial _y\qquad ,{\bf X}_2=\partial _z\qquad ,{\bf X}_3=\partial
_x+y\partial _y-z\partial _z.  \label{sx3.36}
\end{equation}
\begin{equation}
{\bf \omega }^1=e^{-x}dy\qquad ,{\bf \omega }^2=e^xdz\qquad ,{\bf \omega }%
^3=dx.  \label{sx3.36a}
\end{equation}
In this case the non-vanishing structure constants of the isometry group $%
C_{13}^1=-C_{23}^2=1$ and the Jacobi identities (\ref{sx3.1}) imply that the
remaining non-vanishing structure constants $C_{\beta 4}^\alpha $ are:

\begin{equation}
C_{14}^1=a\qquad ,C_{24}^2=b  \label{3.37}
\end{equation}
and the HVF assumes the form: 
\begin{equation}
{\bf H}=\psi t\partial _t+D\partial _x+ay\partial _y+bz\partial _z.
\label{sx3.38}
\end{equation}
Furthermore the four-velocity and the metric are:

\noindent {\bf Fluid velocity}

\begin{equation}
\Delta _1=v_1t^{(\psi +D-a)/\psi }\qquad ,\Delta _2=v_2t^{(\psi -D-b)/\psi
}\qquad ,\Delta _3=v_3t  \label{sx3.39}
\end{equation}
{\bf Metric}

\begin{equation}
g_{\alpha \beta }=\left( 
\begin{array}{ccc}
c_{11}t^{2(\psi +D-a)/\psi } & c_{12}t^{(2\psi -a-b)/\psi } & 
c_{13}t^{(2\psi +D-a)/\psi } \\ 
c_{12}t^{(2\psi -a-b)/\psi } & c_{22}t^{2(\psi -D-b)/\psi } & 
c_{23}t^{(2\psi -D-b)/\psi } \\ 
c_{13}t^{(2\psi +D-a)/\psi } & c_{23}t^{(2\psi -D-b)/\psi } & c_{33}t^2
\end{array}
\right)  \label{sx3.40}
\end{equation}
where $v_\alpha ,c_{\alpha \beta },a,b,D$ are constants. We note that when
the constants $c_{13},c_{23}$ are non-vanishing we may use the freedom $%
y=\pm \tilde{y}/c_{13}$ and $z=\pm \tilde{z}/c_{23}$ to set $%
c_{13}=c_{23}=\pm 1$.

\underline{{\bf Type VII}$_0$}

The KVs $\{X_\alpha \}$ and the dual basis $\{{\bf \omega }^\alpha \}$ are:

\begin{equation}
{\bf X}_1=\partial _y\qquad ,{\bf X}_2=\partial _z\qquad ,{\bf X}_3=\partial
_x-z\partial _y+y\partial _z.  \label{sx3.41}
\end{equation}
\begin{equation}
{\bf \omega }^1=\cos xdy+\sin xdz\qquad ,{\bf \omega }^2=-\sin xdy+\cos
xdz\qquad ,{\bf \omega }^3=dx.  \label{sx3.41a}
\end{equation}
The non-vanishing structure constants are $C_{13}^2=-C_{23}^1=1$ and the
remaining non-vanishing structure constants $C_{\beta 4}^\alpha $ are:  
\begin{equation}
C_{14}^1=C_{24}^2=a.  \label{sx3.42}
\end{equation}
The HVF has the form: 
\begin{equation}
{\bf H}=\psi t\partial _t+D\partial _x+ay\partial _y+az\partial _z.
\label{sx3.43}
\end{equation}
The frame components of the four-velocity and the metric are:

{\bf Fluid velocity}

\[
\Delta _1=t^{(\psi -a)/\psi }\left[ v_1\cos \left( \frac{D\ln t}\psi \right)
+v_2\sin \left( \frac{D\ln t}\psi \right) \right] 
\]
\begin{equation}
\Delta _2=t^{(\psi -a)/\psi }\left[ -v_1\sin \left( \frac{D\ln t}\psi
\right) +v_2\cos \left( \frac{D\ln t}\psi \right) \right]  \label{sx3.44}
\end{equation}
\[
\Delta _3=v_3t 
\]
{\bf Metric}

\begin{equation}
g_{\alpha \beta }=\left( 
\begin{array}{ccc}
t^{2(\psi -a)/\psi }\left[ c_{21}\cos \left( \frac{2D\ln t}\psi \right)
+c_{12}\sin \left( \frac{2D\ln t}\psi \right) +\frac{c_{11}}2\right]  & 
g_{21} & g_{31} \\ 
t^{2(\psi -a)/\psi }\left[ -c_{21}\sin \left( \frac{2D\ln t}\psi \right)
+c_{12}\cos \left( \frac{2D\ln t}\psi \right) \right]  & c_{12}t^{2(\psi
-a)/\psi }-g_{11} & g_{32} \\ 
t^{(2\psi -a)/\psi }\left[ c_{31}\cos \left( \frac{D\ln t}\psi \right)
+c_{13}\sin \left( \frac{D\ln t}\psi \right) \right]  & 
\begin{array}{c}
t^{(2\psi -a)/\psi }[-c_{31}\sin \left( \frac{D\ln t}\psi \right) + \\ 
+c_{13}\cos \left( \frac{D\ln t}\psi \right) ]
\end{array}
& c_{33}t^2
\end{array}
\right)   \label{sx3.45}
\end{equation}
where $v_\alpha ,c_{\alpha \beta },a,D$ are constants.

Applying the translation $x=\tilde{x}+A$ the form of the metric and the KVs
remain invariant. Under this freedom in the choice of the $x-$coordinate we
may set one of the constants $c_{13},c_{31}$, say $c_{31}$, equal to zero
i.e. $c_{31}=0$. The other arbitrary constant, if not vanish, is fixed so
that $c_{13}=\pm 1$ using the constant scale transformations $y=\pm \tilde{y}%
/c_{13}$ and $z=\pm \tilde{z}/c_{13}$.

\section{Tilted perfect fluid solutions}

\setcounter{equation}{0}

Having found the general form of the Bianchi type II, VI$_0$ and VII$_0$
metrics which admit a proper HVF, we may utilize these results to locate
those space-times which have a tilted perfect fluid as their source. In this
case the FE (\ref{sx1.1}) reduced to a purely, although highly non-linear,
algebraic system of equations. However using the restrictions on the frame
components of the fluid velocity described in sections 2,3 we have been able
to solve completely the system of equations for Bianchi type II and VII$_0$
models and present some useful results concerning the Bianchi VI$_0$ models.

The outline of the method is as follows:

For each self-similar Bianchi metric we compute the Ricci tensor $R_{ab}$
and using the decomposition (\ref{sx2.11}) we construct the tensor $%
R_{ab}-\gamma \tilde{\mu}u_au_b-\frac{(2-\gamma )}2\tilde{\mu}g_{ab}$ which,
together with (\ref{sx2.14}), lead to 11 algebraic equations in 16 unknowns $%
c_{\alpha \beta },a,b,\Gamma ,v_\alpha ,\gamma ,\tilde{\mu},D,\psi $. We
note that the number of the essential parameters needed for the general
solution is reduced considerable for each Bianchi model as we shall see in
the analysis presented below.

\underline{{\bf Type II}}

Tilted perfect fluid Bianchi type II models has necessarily zero rotation 
\cite{King-Ellis}. By virtue of equation (\ref{sx2.31}) it turns out that $%
\Delta _1=0$. Furthermore, taking into consideration (\ref{sx2.13}), the
frame component $\Delta _2$ can be set so that $\Delta _2=0$. Therefore the 
{\em general form} of the tilted velocity in self-similar Bianchi II models
is: 
\begin{equation}
u_a=\Gamma \left( n_a+\Delta _3\omega _a^3\right) .  \label{sx4.1}
\end{equation}
Analytical computations show that only in case A$_1$ a perfect fluid
solution exist which subsequently is the most {\em general self-similar
tilted perfect fluid Bianchi II solution}. The metric is given by (\ref
{sx3.26}) with $b=c_{13}=c_{23}=0$, $c_{12}=1$ and the remaining
non-vanishing constants are:

\begin{equation}
c_{11}=\frac{2p(3-4p)}{5p-1}\quad ,c_{22}=\frac{(1-4p)}{2p(5p-3)}\quad
,c_{33}=\frac{(4p-3)(5p-3)}{(4p-1)(5p-1)}  \label{sx4.2a}
\end{equation}
where $p$ is constant and we have set $\psi=\frac 1{1-3p},a =\frac{p-1}{3p-1}
$.

The kinematical and dynamical quantities of this class of self-similar
tilted perfect fluid models are: 
\begin{equation}
\tilde{\mu}=\frac{2p(2p+1)}{t^2},\quad \gamma =\frac 2{2p+1},\quad \Gamma =%
\frac{\sqrt{2}(3-4p)(5p-3)}{2\sqrt{(5p-3)(3-4p)(9p-4)}}  \label{sx4.2b}
\end{equation}
where without loss of generality $v_3=1$.

Taking into account the overall signature of the space-time, the positivity
of the quantities (\ref{sx4.2b}) show that these models are defined for $%
p\in (\frac 35,\frac 34)$ {\em or} $p\in (0,\frac 15)$. The first set
represents an exact Bianchi II perfect fluid solution where the parameter $%
\gamma $ lies in the interval $[4/5,10/11]$. However this solution is not
physically acceptable because the magnitude of the shear tensor $\tilde{%
\sigma}_{ab}\tilde{\sigma}^{ab}$ and the measure of the acceleration $\dot{u}%
^a$ are both negative. In the second set of solutions the state parameter $%
\gamma \in [10/7,2]$ and has been found by Hewitt \cite
{Hewitt,Hewitt-Bridson-Wainwright}.

\underline{{\bf Types VI}$_0$, {\bf VII}$_0$}

In contrast with the Bianchi type II models where significant information
are available, little work has been done for tilted perfect fluid Bianchi VI$%
_0$ and VII$_0$ models even in the special case of self-similar models, due
to additional degrees of freedom in the determination of the general
solution. The problem is simplified with the help of equations (\ref{sx2.31}%
) and (\ref{sx2.32}) which, in conjunction with the FE and the generic form
of the self-similar metrics (\ref{sx3.40}) and (\ref{sx3.45}), can be used
to make some interesting comments concerning the known Bianchi VI$_0$
solutions. As far as we know the only known tilted perfect fluid solution is
due to Rosquist and Jantzen \cite{Rosquist1,Rosquist-Jantzen1}. These
Bianchi VI$_0$ models have non-zero vorticity and the parameter $\gamma \in
(1.0411,1.7169)$ ($\gamma \neq 10/9$). In the case of type VII$_0$ we show
that {\em there are no non-vacuum self-similar solutions}.

From the above discussion one may wonder if there are non-rotating
self-similar Bianchi VI$_0$ and VII$_0$ tilted perfect fluid models at all.
Although the isometry group $G_3$ of Bianchi VI$_0$ and VII$_0$ models,
shares the common property of all the Bianchi models (except types VIII and
IX) to admit an Abelian $G_2$ subgroup, self-similar models are {\em %
necessarily rotating}. In fact we prove the following:

{\em Proposition 2}

\noindent {\it There are no self-similar and irrotational tilted perfect
fluid Bianchi type VI}$_0${\it \ and VII}$_0${\it \ models. }

{\em Proof}

\noindent Projecting equations ${\cal L}_{{\bf X}_\alpha }u_a=0$ and ${\cal L%
}_{{\bf H}}u_a=\psi u_a$ with ${\bf H}$ and ${\bf X}_\alpha $ and
substracting we obtain:

\begin{equation}
2u_{[a;b]}H^aX_\alpha ^b=-\psi u_kX_\alpha ^k+u_kC_{\alpha 4}^\beta X_\beta
^k  \label{sx4.4}
\end{equation}
which relates the projections of $u_{[a;b]}$ in the directions of the HVF
and the KVs, with the structure of the homothetic algebra. Moreover
equations (\ref{sx2.23}) and (\ref{sx2.24}) imply:

\begin{equation}
\dot{u}_{[a}u_{b]}H^aX_\alpha ^b=\psi \frac{\gamma -1}\gamma u_kX_\alpha ^k.
\label{sx4.5}
\end{equation}
By means of equation (\ref{sx2.31}), the condition $\tilde{\Omega}_{ab}=0$
implies that the frame components $\Delta _2$ and $\Delta _3$ vanish
therefore the KVs ${\bf X}_1,{\bf X}_2$ are normal to the tilted velocity
i.e. $X_1^au_a=X_2^au_a=0$. Since $u_{[a;b]}=-\dot{u}_{[a}u_{b]}$ from (\ref
{sx4.4}) and (\ref{sx4.5}) it follows that $\gamma =2$ (we recall that $%
C_{34}^\alpha =0$) i.e. the only transitively self-similar Bianchi VI$_0$
and VII$_0$ tilted perfect fluid models contain stiff fluid as its source.
Then the FE (\ref{sx2.11}), imply that the Ricci tensor has two eigenvectors 
$X_1^a,X_2^a$ with zero eigenvalue: 
\begin{equation}
R_{ab}X_1^b=R_{ab}X_2^b=0.  \label{sx4.6}
\end{equation}
Hence the KVs ${\bf X}_1,{\bf X}_2$ must be hypersurface orthogonal \cite
{Kundt-Trumber} and the metrics are necessarily diagonal. Taking the metrics
(\ref{sx3.40}) and (\ref{sx3.45}) to be diagonal i.e. $%
c_{12}=c_{21}=c_{13}=c_{31}=c_{23}=0$ a self-similar solution occurs only in
type VI$_0$ (the type VII$_0$ solution corresponds to the flat
Friedmann-Lema\^{i}tre model). This solution is found to be unphysical ($%
\tilde{\mu}<0$) in agreement with the general result that the {\em only
non-vacuum self-similar Bianchi model with an Abelian subgroup of isometries
acting orthogonally transitively is given by the metric (\ref{sx3.26}) and (%
\ref{sx4.2a})} \cite{Hewitt}.

In the case of rotating models our analysis shown that when $\gamma =2$ the
resulting type VI$_0$ model is unphysical ($\tilde{\mu}<0$) hence there are
no, physically acceptable, self-similar stiff fluid Bianchi VI$_0$
solutions. We have been unable to solve the resulting system of algebraic
equations in full generality. However, although the family of solutions
found in \cite{Rosquist-Jantzen1} belongs to the special classes satisfying
the constraint $n_\alpha ^\alpha =0$, we expect that this family represents
the most {\em general non-vacuum Bianchi VI}$_0${\em \ model}. In fact
setting $a=(p_1+p_2)\psi +4\psi -b$ and $D=(2-p_2)-b$ in (\ref{sx3.40}) ($%
p_1,p_2$ are constants) it follows that the number of real degrees of
freedom in a self-similar models is one (eleven equations and twelve
unknowns). This issue is currently under investigation.

Concerning the type VII$_0$ models, the method described earlier in this
section does not apply due to the complexity of the expression of the Ricci
tensor. An alternative method to overcome the difficulties of computing the
Ricci tensor, is the use of the Ricci identities for the timelike vector
field $n^a$ combined with equation (\ref{sx2.32}) and has as follows:

The first set of equations we will need is the constraint equations for the
timelike congruence $n^a$ follow from the Ricci identities $%
R_{akbc}n^k=2n_{a[;cb]}$ (equation (72) of \cite{Ellis1}):

\begin{equation}
h_c^ah_d^b\left( \sigma ^{cd}\right) _{;b}+q^a=0.  \label{sx4.7}
\end{equation}
Contracting (\ref{sx4.7}) with the KVs $X_\alpha ^a$ and using the first of (%
\ref{sx2.19}) we obtain: 
\begin{equation}
\left( \sigma _d^aX_\alpha ^d\right) _{;a}+\Gamma \gamma \tilde{\mu}%
u_aX_\alpha ^a=0.  \label{sx4.8}
\end{equation}
In addition expressing equation (\ref{sx2.32}) in terms of the KVs\footnote{%
We can always write the invariant vector fields as combination of the KVs in
the form ${\bf e}_\alpha =Z_\alpha ^\beta {\bf X}_\beta $ where $Z_\alpha
^\beta $ are time-independent smooth functions.} we get: 
\begin{equation}
\dot{u}_{[a}u_{b]}X_\alpha ^aX_\beta ^b=0.  \label{sx4.9}
\end{equation}
The set of equations (\ref{sx4.8})\ and (\ref{sx4.9}) is used to examine the
consistency of the self-similar metrics (\ref{sx3.45}) with the assumption
of a tilted perfect fluid source. Because the expressions of the KVs and
tilted velocity are given by (\ref{sx3.41}) and (\ref{sx3.44}) respectively,
we compute the quantity $\sigma _d^aX_\alpha ^d$ and the acceleration $\dot{u%
}_a$ of the fluid. Replacing back to (\ref{sx4.8})\ and (\ref{sx4.9}) we are
left with a system of 6 algebraic equations whose solution gives the
required answer. It is found that for every choice of the parameters, the
energy density is always negative except in the case where the space-time
reduced to the flat Robertson-Walker space-time. Hence {\em there are no
self-similar and tilted perfect fluid Bianchi type VII}$_0${\em \ model
except its flat RW counterpart}.

\section{Discussion}

In this paper we have been able to give the explicit forms of the non-vacuum
self-similar Bianchi models of class A regardless the matter content of the
gravitational field. Therefore the results found in section 3, although
purely geometrical, can be used to determine the general tilted perfect
fluid self-similar solution since the only limited kinematical assumption we
have made is the conformally mapping of the fluid velocity i.e. ${\cal L}_{%
{\bf H}}u^a=-\psi u^a$. It follows that fluid models (non-tilted or tilted)
with arbitrary energy-momentum tensor which admit a proper HVF, must
included to the general models presented here.

An important aspect of the analysis given in sections 3 and 4 (which is
valid for the entire class of Bianchi models) is the direct dependency of
the vorticity on the structure of the similarity group. This observation
enables us to make some straightforward comments regarding the behaviour of
the non-vacuum Bianchi models class B. In particular from equations (\ref
{sx2.31}), (\ref{sx4.4}) and (\ref{sx4.5}) we deduce that for {\em all}
irrotational Bianchi models, except type III, the fluid velocity frame
components $\Delta _1=\Delta _2=0$. Thus we expect that self-similar class B
models subject to similar restrictions as in class A models. In fact it can
be shown that self-similar and irrotational Bianchi IV and V models are only
those for which $\gamma =2$ \cite{Apostol8}. It turns out that the
corresponding KVs ${\bf X}_1,{\bf X}_2$ are also hypersurface orthogonal and
the metric can be written in diagonal form in which case neither solution is
physically acceptable hence {\em there are no Bianchi IV and V self-similar
tilted perfect fluid solutions with zero vorticity}.

Concerning the rotational class A models one must to solve the high
non-linear algebraic equations in order to answer definitely whether or not
a self-similar perfect fluid model exists. We have been able to work
efficiently on the case $\gamma =2$ in Bianchi type VI$_0$ and show that
this class of models, found by Wainwright et al. \cite
{Wainwright-Ince-Marshman}, do not admit a proper HVF. In addition based on
the fact that there is only one essential parameter together with a
rigorous, but not conclusive, study of the self-similar-metrics (\ref{sx3.40}%
),\ we regard the assumption that the family of Rosquist-Jantzen solutions
is the most general in type VI$_0$, as highly plausible.

In the case of Bianchi type VII$_0$ our analysis shows that {\em %
self-similar models do not admit a perfect fluid interpretation}. As we have
mentioned in the Introduction types VIII and IX also do not admit a perfect
fluid interpretation and it is well known that the corresponding non-tilted
models share the same feature (see e.g. Table 1 of \cite{Hsu-Wainwright}).
These results indicate that, in connection with the {\em self-similarity
breaking at late times} in Bianchi VII$_0$ non-tilted models \cite
{Wainwright-Hancock-Uggla,Horwood-Hancock-The-Wainwright}, it may be
possible to relate the existence of a self-similar Bianchi model with the
asymptotic self-similarity property of more general models. Consequently, it
is an interesting and important task to determine under what conditions a 
{\em full self-similarity breaking} (i.e. in the asymptotic regimes) occurs
and prove the conjecture that a {\em general perfect fluid Bianchi model is
asymptotically self-similar at late times or near to the initial singularity
if and only if a self-similar perfect fluid model exists}.

\end{document}